\documentstyle[twocolumn]{jpsj}
\def\runtitle{Non-magnetic Impurities in Spin Gap Systems}
\def\runauthor{N.~{\sc Nagaosa}, A.~{\sc Furusaki}, M.~{\sc Sigrist}
 and H.~{\sc Fukuyama}} 

\title{Non-magnetic Impurities in Spin Gap Systems}

\author{
Naoto {\sc Nagaosa}, Akira {\sc Furusaki},$^1$
Manfred {\sc Sigrist}$^{1,}$\footnote{Present address: Theoretische
 Physik, ETH-H\"onggerberg, 8093 Z\"urich, Switzerland.}
and Hidetoshi {\sc Fukuyama}$^2$}

\inst{
Department of Applied Physics, University of Tokyo,
Bunkyo-ku, Tokyo 113 \\
$^1$Yukawa Institute for Theoretical Physics, Kyoto University,
Kyoto 606-01 \\
$^2$Department of Physics, University of Tokyo,
Bunkyo-ku, Tokyo 113}
\recdate{August 27, 1996}

\abst{
The effect of non-magnetic impurities is discussed for both the two-leg
Heisenberg ladder system and other spin liquids with excitation gap.
It is shown that the random depletion of spins introduces a random
Berry phase term to the non-linear $\sigma$ model.
The classical nature of the antiferromagnetic correlation is enhanced
by the topological decoherence, and the staggered susceptibility shows
more singular behavior at low temperatures than the uniform
antiferromagnetic Heisenberg chain.
}
\kword{quantum spin chain, Heisenberg ladder, random spin system}

\begin{document}
\sloppy
\maketitle

Recently a class of quantum antiferromagnets which has a ground state 
with a gap in the spin excitation spectrum has attracted much
attention.
These systems are characterized either by a local singlet correlation
of valence-bond-solid state as found in spin-Peierls systems\cite{HASE}
and even-leg spin ladders\cite{DAGOTTO} or by resonating-valence-bond
(RVB) state as in underdoped cuprates.
In such systems it was found that the presence of non-magnetic
impurities replacing magnetic ions often leads to a Curie behavior of
the uniform susceptibility with a Curie constant
proportional to the impurity concentration.
This can be understood basically from the idea
that, due to the depletion of spins, some spins lose their partners
to form a singlet.
They appear, therefore, as nearly independent spin doublet degrees of
freedom.
However, the more striking experimental feature is the appearance of
the antiferromagnetic long-range order (AFLRO) induced by the
non-magnetic impurities.
This property has been observed essentially in all the systems
mentioned above,\cite{EXP1,EXP2,EXP3} and thus seems to be a universal 
phenomenon in gapped quantum spin liquids.
This effect has recently been analyzed in terms of the phase
Hamiltonian for the spin-Peierls system by Fukuyama {\it et
al.}\cite{FUKU1}
The presence of impurities leads to a local weakening of the
dimerization, which induces enhanced local antiferromagnetic
correlations.
The AFLRO can then emerge at low temperatures, because the local
antiferromagnetic moments around the impurities correlate in phase
with each other.
The phase Hamiltonian approach has also been applied to the two-leg
spin ladder system with a single non-magnetic impurity.\cite{FUKU2}
In this case the phase field forms a soliton at the impurity position,
which corresponds to an impurity-induced spin degree of freedom ($S=1/2$).
Local antiferromagnetic correlation appears in the vicinity of
the impurity, and would turn to an in-phase (quasi) long-range ordered
state in the ground state.\cite{FUKU2}
Sigrist and Furusaki pointed out that low-energy physics of the lightly
depleted spin ladder system is described by a spin-$\frac{1}{2}$
Heisenberg model with random ferromagnetic and antiferromagnetic
couplings.\cite{SIFU}
They found that at low temperatures the uniform susceptibility
follows the Curie law while the specific heat shows power-law
behavior, $C\sim T^{2\alpha}$, where the exponent $\alpha$ is expected 
to be equal to or smaller than 0.25.
On the other hand, in a recent numerical study Furukawa {\it et al.}\
found a significant enhancement of the staggered susceptibility
$\chi(Q)$ in the presence of non-magnetic impurities.\cite{FURU}
The asymptotic behavior, however, could not be obtained because the
size of the system they studied is not large enough.

In this paper we study the effect of non-magnetic impurities on gapped 
spin liquids via a mapping to the non-linear $\sigma$ model with a
random topological term.
In this formulation the appearance of the AFLRO originates from the
recovery of the topological Berry phase term, which suppresses quantum 
fluctuations such that the classical nature of the AFLRO is enhanced.
We find that the generalized susceptibility
$\chi(Q,T)$ diverges with a non-trivial exponent,
$\chi(Q,T) \sim T^{-1-2\alpha}$,
in the low-temperature limit.
This divergence is much stronger than the spin-$\frac{1}{2}$
antiferromagnetic Heisenberg chain.
This picture also applies to higher-dimensional systems
where quantum fluctuations suppress the AFLRO and generate the spin
gap phase, even though the behavior of
$\chi(Q,T)$ depends on the dimensionality.

For concreteness we consider the two-leg spin-$S$ Heisenberg ladder as
an example of gapped spin liquids.
We use the functional integral method and write the partition function
$Z$ as
\begin{equation}
Z = \int D {\mib S}_i(\tau) e^{- A}.
\end{equation}
The action $A$ of the system is given by
\begin{equation}
A = {\rm i} S \sum_i \omega(\{ {\mib S}_i(\tau) \})
+ \int {\rm d}\tau \sum_{<ij>} J {\mib S}_i(\tau) \cdot {\mib S}_j(\tau). 
\end{equation}
The first term is the Berry phase contribution, where
$\omega(\{ {\mib S}_i(\tau) \})$
is the solid angle enclosed by 
the closed trajectory of ${\mib S}_i(\tau)$ [${\mib S}_i(0)={\mib
S}_i(\beta)$].\cite{FRADKIN}
The second term represents the nearest-neighbor antiferromagnetic
interaction.

For the following discussion it is crucial to notice that no
frustration is induced by the random depletion of spins because the
bipartite lattice structure remains and the antiferromagnetic
interaction works only between $A$- and $B$-sublattice spins.
Consequently we should choose the staggered spin
as the slowly varying field.
On the other hand, the Hamiltonian commutes with the total spin
so that the Fourier components with small wavevectors are also expected
to be slowly varying.
From these considerations we may express a spin
operator in terms of two slowly varying fields,
${\mib \Omega}( x_i, \tau)$
  and
${\mib \Pi}( x_i, \tau)$:
\begin{equation}
{\mib S}_i(\tau) =
 S[ (-1)^i {\mib\Omega}(x_i,\tau) + a {\mib\Pi}(x_i,\tau)].
\end{equation}
Here $a$ is the lattice constant, and the two fields satisfy the
constraint, ${\mib\Omega}\cdot{\mib\Pi} = 0$ and $|{\mib\Omega}|=1$.
We can then apply the standard procedure to derive the non-linear
$\sigma$ model.

Before discussing the randomly depleted Heisenberg ladder we first
consider pure spin systems.
It is well known that the effective action $A$ for a single Heisenberg 
chain is\cite{FRADKIN}
\begin{equation}
A = 
A_{\rm Berry}
 + \frac{1}{2g} \int_0^{\beta}\!{\rm d}\tau \int\!{\rm d}x
\left[\frac{1}{c}
\left(\frac{\partial{\mib\Omega}}{\partial \tau}\right)^2
+ c \left(\frac{\partial{\mib\Omega}}{\partial x}\right)^2 \right],
\label{eq:action}
\end{equation}
where the dimensionless coupling constant $g = 2/S$
controls the quantum  fluctuations, and the velocity $c = 2JSa $ sets
the energy scale.
The Berry phase term $A_{\rm Berry}$ is given by
\begin{equation}
A_{\rm Berry} = 2 \pi {\rm i} Q S ,
\end{equation}
where $Q$ is an integer called Skyrmion number,
\begin{equation}
Q = \frac{1}{4 \pi} \int_0^{\beta} {\rm d}\tau \int{\rm d}x\,
{\mib\Omega} \!\cdot\! \left( 
\frac{\partial{\mib\Omega}}{\partial \tau}
\times
\frac{\partial{\mib\Omega}}{\partial x}
\right).
\end{equation}
For half integer $S$ the Berry phase is the multiple of
$\pi$ and thus causes destructive interference between the quantum 
fluctuations with different Skyrmion numbers.
This leads to the massless spectrum of the triplet excitations.
For integer $S$, however, the Berry phase plays no role for the
infinite system and quantum fluctuations generate the spin gap.\cite{HALDANE}
In the case of the pure two-leg spin ladder, one leg of the ladder
contributes $A_{\rm Berry}$ to the effective action, while
the other leg gives $A_{\rm Berry}({\mib\Omega}\to-{\mib\Omega})$.
The Berry phase terms thus cancel, and the action is just the
non-linear $\sigma$ model without a topological term.\cite{SENE,SIERRA}
Hence the spin gap opens similarly to the integer spin chain.

Now we turn to the randomly depleted spin ladder.
We note that removing spins from the ladder affects the effective
action in two ways.
An obvious effect is the local weakening of the coupling $J$
which leads to the reduction in the sound velocity $c$.
This tends to decrease the energy scale of the problem.
The other effect is of quantum mechanical nature and is related to the 
Berry phase.
The cancelation of the Berry phases as described above is violated by
the depleted spins, and the Berry phase term $A_{\rm Berry}$ becomes
\begin{eqnarray}
A_{\rm Berry} 
&=& {\rm i} S \sum_{i \ne i_n} 
\omega(\{ {\mib S}(x_i,\tau) \})
\nonumber \\
&=& {\rm i} S \sum_{i} 
\omega(\{ {\mib S}(x_i,\tau) \})
- {\rm i} S \sum_{n} 
\omega(\{ {\mib S}(x_{i_n},\tau) \})
\nonumber \\
& = &  - {\rm i} S \sum_{n} (-1)^{i_n}
\omega(\{ {\mib\Omega}(x_{i_n},\tau) \}),
\label{eq:A_Berry}
\end{eqnarray}
where $i_n$ specifies the site of a depleted spin.
This modification leads to a substantial change in the quantum
mechanical behavior of the system described by eqs.~(\ref{eq:action})
and (\ref{eq:A_Berry}), as we will see later.

In the next step we integrate over the continuum degrees of freedom
${\mib\Omega}(x\ne x_{i_n},\tau)$ 
to derive the effective action for
${\mib I}_n(\tau) \equiv {\mib\Omega}(x= x_{i_n},\tau)$.
To this end we introduce the Lagrange multiplier ${\mib\lambda}_i$
to implement the constraint
${\mib I}_n(\tau) = {\mib\Omega}(x= x_{i_n},\tau)$, and write $Z$ as
\begin{equation}
Z = \int D {\mib\Omega} \int D {\mib\lambda}
    \int D {\mib I} \, e^{- A_{\rm eff}}.
\end{equation}
The effective action $A_{\rm eff}$ is given by
\begin{eqnarray}
A_{\rm eff} &=&
- {\rm i} S \sum_n (-1)^{i_n} \omega( \{ {\mib  I}_n(\tau) \} )
\nonumber \\
&&
+ \frac{c}{2g}\sum_{k, \omega_l}
  \!\left[\left(\frac{\omega_l}{c}\right)^2\!+k^2+\frac{1}{l_0^2}\right]\! 
  {\mib\Omega}(k,\omega_l) \!\cdot\! {\mib\Omega}(-k,-\omega_l) 
\nonumber \\
&&
+ {\rm i} \sum_n \sum_{\omega_l}
{\mib\lambda}_n(- \omega_l) \!\cdot\!\!
\left(\!
{\mib I}_n(\omega_l)\!-\!
\sum_k\frac{{\rm e}^{{\rm i}kx_{i_n}}}{\sqrt{L}}{\mib\Omega}(k,\omega_l)
\!\right)\!\! , \nonumber\\&&
\label{eq:massive}
\end{eqnarray}
where $\omega_l$ is the Matsubara frequency for bosonic fields, and
$(-1)^{i_n}=+1$ $(-1)$ if $x_{i_n}$ is on the $A$-sublattice
($B$-sublattice).
In eq.~(\ref{eq:massive}) we have introduced a mass $c/l_0$ in the
second term.
This corresponds to taking saddle-point approximation for
a Lagrange multiplier field representing the constraint
$|{\mib\Omega}|=1$.
This can be justified, in the limit of small
impurity concentration, by noting that $l_0$ is the finite
correlation length of the pure two-leg spin ladder.
We then integrate over ${\mib\Omega}$ to obtain
\begin{eqnarray}
A_{\rm eff} &=& 
- {\rm i} S \sum_n (-1)^{i_n} \omega(\{ {\mib  I}_n(\tau) \})
\nonumber \\ &&
+ {\rm i} \sum_n \sum_{\omega_l}
   {\mib\lambda}_n(- \omega_l) \cdot  {\mib I}_n(\omega_l)
\nonumber \\ & &
+ \frac{g}{2c} \sum_{m,n} K_{m,n}(\omega_l) 
{\mib\lambda}_m( \omega_l) \cdot {\mib\lambda}_n(- \omega_l), 
\end{eqnarray}
where we have ignored an additive constant coming from integration
over ${\mib\Omega}$, and
\begin{equation}
K_{m,n}(\omega_l) = \int^{\infty}_{-\infty} \frac{{\rm d}k}{2\pi}
\frac{{\rm e}^{{\rm i} k (x_{i_m} - x_{i_n} )}}{k^2 + k_0^2}
=\frac{1}{2k_0} {\rm e}^{-k_0|x_{i_m} - x_{i_n} |}
\end{equation}
with $k_0=\left[(\omega_l/c)^2+l_0^{-2}\right]^{1/2}$.
For the case where the concentration $\delta$ of the 
non-magnetic impurities is very small ($a/\delta\gg l_0$), we can
approximate $K^{-1}$ as
\begin{equation}
(K^{-1})_{m,n}\approx
2k_0\left( \delta_{m,n} - \delta_{m,n \pm 1}
{\rm e}^{ -k_0|x_{i_m} - x_{i_n}| }\right).
\end{equation}
After integrating over ${\mib\lambda}_i$ we obtain,
for $|\omega_n|\ll c/l_0$,
\begin{eqnarray}
A_{\rm eff}&=&
- {\rm i} S \sum_n (-1)^{i_n} \omega(\{ {\mib  I}_n(\tau) \})
\nonumber \\
&&
- \!\int_0^{\beta}\!{\rm d}\tau
\sum_{n} \tilde{J} {\rm e}^{ - |x_{i_n} - x_{i_{n+1}}|/l_0 }
  {\mib I}_n (\tau) \!\cdot\! {\mib I}_{n+1} (\tau), \qquad
\label{eq:effective-action}
\end{eqnarray}
where $\tilde J=ck_0/g$.
Equation (\ref{eq:effective-action}) can be identified with the action
for the random-exchange Heisenberg model (REHM), 
\begin{equation}
H = - \sum_n J_n {\mib S}_n \cdot {\mib S}_{n+1},
\label{eq:REHM}
\end{equation}
where ${\mib S}_n=(-1)^{i_n+1}{\mib I}_n$.
The exchange interaction $J_n$ is given by
$J_n = \tilde{J} (-1)^{i_n + i_{n+1} }
\exp( - |x_{i_n} - x_{i_{n+1}}|/l_0 )$, which is random in sign 
as well as in magnitude.\cite{SIFU}
Note that the two spins ${\mib S}_n$ and ${\mib S}_{n+1}$ are coupled
ferromagnetically (antiferromagnetically) if they are on the same
(different) sublattice(s).\cite{SIFU}
It is important to notice that the dependence of the 
effective action on the impurity concentration $\delta$
appears only in the magnitude of the exchange interaction $J_n$,
which is typically of the order of $J {\rm e}^{- l_0/\delta}$.

The low-energy properties of the REHM have recently been studied by
Westerberg {\it et al.}\ using a real-space renormalization group
method.\cite{WESTER,WESTER2,FURUSAKI}
They found that the correlation length $\xi$ for the staggered spin
${\mib I}_n$ diverges at low temperatures as\cite{WESTER2}
\begin{equation}
\xi \sim T^{-2 \alpha},
\end{equation}
where the exponent $\alpha$ was estimated numerically to be 0.22.
Inside a domain of length $\xi$ the spins ${\mib S}_n$ align almost
parallel or antiparallel to each other depending on the sign of $J_n$.
These strongly correlated spins form a single effective spin, whose spin
size can be estimated from a random-walk argument.\cite{WESTER,FURUSAKI}
Since the sign of $J_n$ is random, the spin size $S_{\rm tot}$ is
given by a random sum of positive and negative numbers, thereby
yielding $S_{\rm tot}\sim\sqrt{\xi}$.
Consequently, the uniform susceptibility $\chi_{\mibs S}(T)$ per site
behaves as
\begin{equation}
\chi_{\mibs S}(T) \sim \frac{1}{T} 
\frac{S_{\rm tot}^2}{\xi(T)} \sim \frac{1}{T},
\end{equation}
and the Curie law persists down to zero temperature.\cite{WESTER}
In contrast to ${\mib S}_n$, the staggered spins ${\mib I}_n$ align
always parallel to each other inside the domain.
Thus the total spin size $I_{\rm tot}$ of the staggered spins
${\mib I}_n$ inside the domain is proportional to $\xi$.
Hence the generalized (staggered) susceptibility $\chi_{\mibs I}(T)$
behaves as
\begin{equation}
\chi_{\mibs I}(T) \sim \frac{1}{T}
\frac{I_{\rm tot}^2}{\xi(T)} \sim \frac{\xi}{T}
\sim T^{-1-2\alpha},
\label{eq:chi_I}
\end{equation}
which is more singular than the staggered susceptibility of the
spin-$\frac{1}{2}$ antiferromagnetic Heisenberg chain.
This implies that the impurity-induced staggered moments
${\mib I}_n$ show at least quasi-long-range order at zero temperature,
which, even with an infinitesimal inter-ladder coupling, will turn
into true long-range order, in agreement with the experimental
observation.

Let us examine the REHM in eqs.~(\ref{eq:effective-action}) and
(\ref{eq:REHM}) in more detail.
As we discussed above, the spin quantum number $S_{\rm eff}$ of the
effective spin representing the correlated spins scales as
$S_{\rm eff}\sim\sqrt{\xi(T)}$, which grows with decreasing temperature.
This can be considered as an intermediate case between the
ferromagnetic Heisenberg chain where $S_{\rm eff} \sim \xi$ and the
antiferromagnetic Heisenberg chain where $S_{\rm eff}$ is at most $S$.
This means that the classical nature of ${\mib I}$ is more
enhanced in the randomly depleted Heisenberg ladder than in the pure
spin-$\frac{1}{2}$ antiferromagnetic Heisenberg chain, as we see in
eq.~(\ref{eq:chi_I}).
\begin{fulltable}
\caption{Universality classes of one-dimensional quantum spin chains
classified in terms of the Berry phase.}
\label{tab:Berry}
\begin{fulltabular}{lllll}
\hline
& class  & Berry phase & specific heat &
 generalized susceptibility \\
\hline
1. & ferromagnets & ${\rm i}S\sum_n\omega(\{ {\mib I}_n(\tau)\})$ &
 $C\sim\sqrt{T}$ & $\chi_{\mibs I}\sim T^{-2}$ \\
2. & random-exchange Heisenberg model &
 ${\rm i}S\sum_n\sigma_n\omega(\{ {\mib I}_n(\tau)\})$ ($\sigma_n=\pm1$)
 & $C\sim T^{2\alpha}$
 & $\chi_{\mibs I}\sim T^{-1-2\alpha}$ \\
3. & antiferromagnets (half-integer $S$) & $2\pi{\rm i}SQ$ ($Q$:
Skyrmion number) & $C\sim T$ & $\chi_{\mibs I}\sim T^{-1}$ \\
4. & antiferromagnets (integer $S$) & $2\pi{\rm i}SQ$ ($Q$: Skyrmion number)
 & $C\sim {\rm e}^{-\Delta/T}$ & $\chi_{\mibs I}\sim {\rm const}$ \\
\hline
\end{fulltabular}
\end{fulltable}
We summarize these considerations in Table I, where four universality
classes of one-dimensional quantum spin chains are classified
according to the degree of quantum fluctuations of the field
${\mib I}$.
The Berry phase term is most effective in the 1st class while it plays
no role in the 4th class, where quantum fluctuations dynamically
generate a mass.
The even-leg spin ladders without impurities also belong to this class.
We see that, as the Berry phase term becomes more and more effective,
the classical nature of ${\mib I}$ emerges, leading to the
enhanced generalized susceptibility $\chi_{\mibs I}(T)$.
It is important to notice that any small amount of non-magnetic
impurities is sufficient to change the universality class of the pure
spin liquids, e.g., the even-leg spin ladders, from the 4th
class to the 2nd class.
This is due to the decoherence by the random Berry phase term.
A similar idea has been discussed recently by Prokof'ev and Stamp for
quantum nanomagnets.\cite{STAMP1,STAMP2}
Here some remarks are in order on the random singlet phase, in which
spins far apart form weakly bound singlet pairs in a random
manner.\cite{FISHER,note}
This phase is realized in the random antiferromagnetic Heisenberg
model where the exchange couplings are all antiferromagnetic but
are random in their magnitude.
In this case the Berry phase term is not affected by the randomness,
and the singlet formation of spins plays an essential role.
We did not include this phase in Table I, because the randomness is
not of topological nature.
Therefore Table I is not intended to exhaust all the 
universality classes of the quantum spin chains. 

Finally, we briefly discuss the two-dimensional case for which
the mapping to the non-linear $\sigma$ model can be used in exactly
the same way as in the one-dimensional case.
For example, in the Heisenberg antiferromagnet on the square lattice, 
Berry phases of each row cancel as in the spin ladder, and thus there 
is no topological term for the slowly varying field $\Omega$.\cite{FRADKIN}
For $S=\frac{1}{2}$ the dimensionless coupling constant $g
\sim 1/S$ is not large enough for the quantum fluctuations to destroy
the AFLRO.
However, as the mobile holes are doped into the antiferromagnet (e.g.,
doped high-Tc cuprates), the AFLRO is rapidly suppressed and eventually
destroyed.
These holes are localized when Zn impurities are doped sufficiently.
In this case both the Zn impurities and the localized holes contribute
to the random Berry phase terms in the non-linear $\sigma$ model.
Then we expect that the antiferromagnetic correlation is enhanced and
the AFLRO is recovered with non-magnetic impurities, which is indeed
observed in a recent experiment on the underdoped cuprates with the
(pseudo) spin gap.\cite{EXP3}
The problem of the coexistence with the mobile holes and Zn impurities 
is left for future study.

Before summarizing the results, we point out that a non-linear
$\sigma$ model with a random topological term has also been discussed
in the context of the localization of a particle in a random magnetic
field.\cite{ZHANG}
It is an interesting future problem to explore further this connection 
between these two apparently different problems, the random-exchange
Heisenberg model and the random flux problem.

In summary, we have studied the effects of non-magnetic impurities
in gapful spin liquids.
We have shown that the depletion of spins induces the random Berry
phases, leading to the destructive interference of quantum fluctuations.
The effective low-energy model for the randomly depleted spin ladder is
the random-exchange Heisenberg chain, in which the generalized
susceptibility for the staggered moment diverges as $T^{-1-2\alpha}$
($\alpha \approx 0.22$).
The correlation of the staggered spin moment becomes more classical
due to the decoherence by the random Berry phase term.

\acknowledgements
We are grateful to N.\ Furukawa, M.\ Imada, T.\ M.\ Rice,
P.\ C.\ E.\ Stamp and S.\ C.\ Zhang for stimulating discussions.
This work was initiated during the workshop on strongly correlated
electron systems at Yukawa Institute for Theoretical Physics, Kyoto
University.
One of the authors (M.S.) is also supported by a fellowship (PROFIL)
of the Swiss Nationalfonds.

\end{document}